\documentclass[conference,9pt]{IEEEtran}
\IEEEoverridecommandlockouts
\usepackage{amsmath,amssymb,amsfonts}
\usepackage{algorithmic}
\usepackage{mathrsfs}
\usepackage{graphicx}
\usepackage{textcomp}
\usepackage{xcolor}
\usepackage{multirow}
\usepackage{booktabs}
\usepackage{makecell}
\usepackage{afterpage}
\usepackage[numbers,sort&compress]{natbib}
\usepackage{xcolor}
\usepackage[hidelinks]{hyperref}

\begin{document}

\title{LMFCA-Net: A Lightweight Model for Multi-Channel Speech Enhancement with Efficient Narrow-Band and Cross-Band Attention
}

\name{Yaokai Zhang$^{\star}$ \qquad Hanchen Pei$^{\star}$ \qquad Wanqi Wang$^{\dagger}$ \qquad Gongping Huang$^{\star}$}

\address{$^{\star}$ School of Electronic Information, Wuhan University, 430072, Wuhan, China \\
$^{\dagger}$ School of Information Management, Wuhan University, 430072, Wuhan, China}

\maketitle
	
	\begin{abstract}
		Deep learning based end-to-end multi-channel speech enhancement methods have achieved impressive performance by leveraging sub-band, cross-band, and spatial information. However, these methods often demand substantial computational resources, limiting their practicality on terminal devices. This paper presents a lightweight multi-channel speech enhancement network with decoupled fully connected attention (LMFCA-Net). The proposed LMFCA-Net introduces time-axis decoupled fully-connected attention (T-FCA) and frequency-axis decoupled fully-connected attention (F-FCA) mechanisms to effectively capture long-range narrow-band and cross-band information without recurrent units. Experimental results show that LMFCA-Net performs comparably to state-of-the-art methods while significantly reducing computational complexity and latency, making it a promising solution for practical applications.
	\end{abstract}


	\section{Introduction}
	\label{sec:intro}
	
	Speech enhancement aims to improve the clarity and intelligibility of speech by suppressing noise and reverberation. It is widely employed in applications such as teleconferencing, hearing aids, and hands-free human-machine interfaces~\cite{benesty2008microphone, michelsanti2021overview}. Nowadays, most systems are equipped with microphone arrays for sound signal acquisition, and the strength of multi-channel speech enhancement lies in its ability to exploit spatial information~\cite{emmanuel2014spatial, huang2018insights}. Traditional approaches are primarily based on signal processing, with common methods including beamforming and multi-channel filtering~\cite{habets2010new, brandstein2013microphone, benesty2023microphone, huang2022fundamental}.

	Recent advances in deep learning have led to a shift in microphone arrays speech enhancement towards deep learning-based approaches~\cite{heymann2016neural}. One strategy involves estimating beamforming parameters using deep neural networks (DNNs)~\cite{heymann2016neural, zhang2021adl}, while end-to-end multi-channel speech enhancement directly processes input speech signals using DNNs to leverage their nonlinear capabilities. To effectively exploit spatial information, end-to-end multi-channel speech enhancement methods often include modules that explicitly capture narrow-band and cross-band information. For instance, the FT-JNF employs long short-term memory (LSTM) units along the time and frequency axes of the spectrogram~\cite{tesch2022insights}, while McNet~\cite{yang2023MCNET} improves this approach by applying LSTM units across combined adjacent frames or frequency bins to analyze spectral information. More recently, the TF-GridNet~\cite{wang2023tf} and SpatialNet~\cite{quan2024spatialnet} have incorporated self-attention mechanisms to model narrow-band and cross-band information. This two-stage approach is advantageous because spatial cues remain consistent over time for stationary sources, though they tend to vary with frequency~\cite{wang2018combining}. Cross-band information, in turn, enhances the integration of these features~\cite{wang2023tf}.
	However, current DNN-based multi-channel speech enhancement methods are often computationally intensive, making them challenging to implement in terminal applications with limited processing capabilities~\cite{michelsanti2021overview}. The sequential nature of recurrent units, such as LSTMs, limits parallelization and hardware efficiency~\cite{vaswani2017attention}, while self-attention mechanisms impose a substantial computational burden due to extensive matrix multiplications~\cite{tang2022ghostnetv2}. Existing lightweight speech enhancement methods primarily focus on mono-channel designs. For instance, PerceptNet~\cite{valin2020perceptually} uses GRU recurrence as the DNN backbone, DeepFilterNet~\cite{schroter2022deepfilternet} and GTCRN~\cite{rong2024gtcrn} reduce computational complexity by employing grouped GRUs. However, these methods still face issues such as gradient vanishing and training instability~\cite{jozefowicz2015empirical}.

    	Implementing an end-to-end multi-channel speech enhancement network on terminal devices is worth exploring. According to previous studies~\cite{stoller2018wave}, simply extending single-channel speech input to multi-channel input and allowing the network to implicitly utilize spatial information can improve the speech enhancement performance.
	Currently, multi-channel speech enhancement networks, such as TF-GridNet~\cite{wang2023tf} and many others \cite{yang2023MCNET, luo2020end} typically have a computational cost of several to tens of billions of multiply-accumulate operations per second (MACs). These complex model structures achieve superior enhancement performance, but they are difficult to practically apply to terminal devices. We summarize the current dilemma in the following two points: 
	
	\begin{itemize}
		\item Only a few multi-channel designs consider lightweight features, for example, NICE-Beam~\cite{casebeer2021nice} estimates statistical quantities within the beamforming framework, focusing primarily on effectively estimating the covariance matrix for minimum variance distortionless response (MVDR) filters.
		\item Modeling narrow-band and cross-band information is essential for enhancing the use of spatial cues~\cite{wang2023tf}, a technique widely used in multi-channel designs~\cite{yang2023MCNET, quan2024spatialnet, wang2018combining}. Previous methods commonly rely on recurrent units or self-attention module to achieve this, which limit computational efficiency and, in the case of recurrent units, can result in training instability~\cite{vaswani2017attention, jozefowicz2015empirical}.
	\end{itemize}

	In this work, we propose an approach that leverages long-range narrow-band and cross-band information without relying on recurrent units. This is achieved through the introduction of time-axis decoupled fully-connected attention (T-FCA) and frequency-axis decoupled fully-connected attention (F-FCA) mechanisms, inspired by the work of Tang et al.~\cite{tang2022ghostnetv2}. Building on these mechanisms, we introduce the \textbf{L}ightweight \textbf{M}ulti-channel speech enhancement network with decoupled \textbf{F}ully \textbf{C}onnected \textbf{A}ttention, termed as LMFCA-Net, to address these challenges. Experimental results demonstrate that the two-stage modeling, based on T-FCA and F-FCA, is highly effective. Furthermore, LMFCA-Net strikes an excellent balance between computational efficiency and enhancement performance.
	
	\section{Problem Description}
		
	\begin{figure*}[t]
		\centering
		\includegraphics[width=\textwidth]{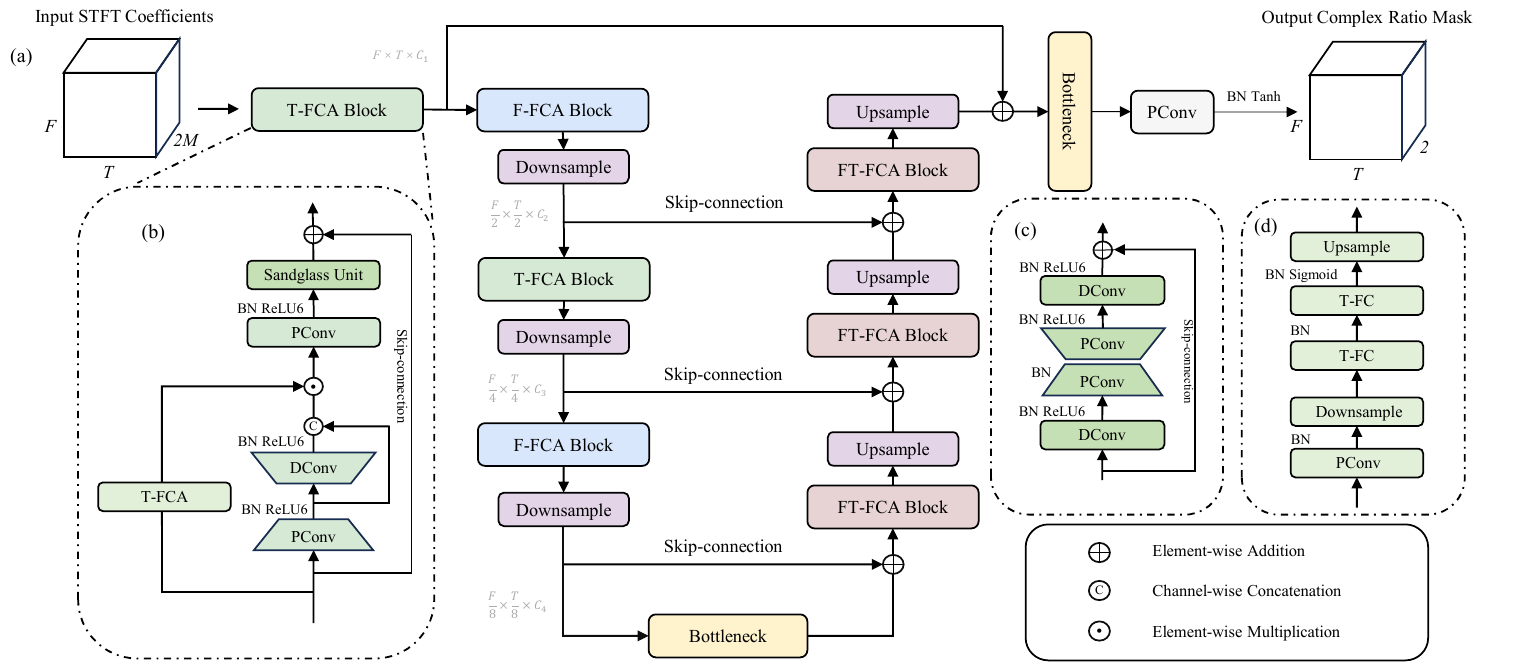}
		\caption{Network architecture: (a) overview of the proposed LMFCA-Net, (b) T-FCA Block, (c) Sandglass Unit, and (d) T-FCA module.}
		\label{fig:Figure 1.}
	\end{figure*}
	\label{sec:PD}
	We consider a signal model where a microphone array consisting of $M$ sensors captures a convolved source signal within a noisy environment. The signal received by the $m$th microphone, denoted as $x_m(t)$ at discrete time index $t$, is expressed as:
	\begin{align}
		\label{y-m(t)}
		x_m(t) &= h_m(t) * s(t) + v_m(t), \ m=1,2,\ldots,M,
	\end{align}
	where $h_m(t)$ represents the acoustic impulse response from the unknown speech source, $s(t)$, to the $m$th microphone, $*$ denotes the linear convolution, and $v_m(t)$ is the additive noise at microphone $m$. The objective of speech enhancement is to recover the target signal $s(t)$ from these microphone observations.
	In the short-time Fourier transform (STFT) domain, the observed signal can be expressed as:
	\begin{align}
		X_{m}(t,f) = \Phi_{m}(t, f) + V_{m}(t,f), \ m=1,2,\ldots,M,
	\end{align}
	where $X_{m}(t,f)$, $\Phi_{m}(t, f)$, $S(t,f)$, and $V_{m}(t,f)$ are the STFTs of $x_m(t)$, $h_m(t) * s(t)$, and $v_m(t)$, respectively, with $t$ being the discrete time index. We conduct speech enhancement in the STFT domain.

	\section{Proposed Method}
	\label{sec:PM}
	
	To address the aforementioned challenges, we propose LMFCA-Net with the FCA mechanisms. The advantages of LMFCA-Net are two folds:
	It is significantly more efficient than state-of-the-art methods while maintaining comparable performance, as demonstrated by its lower GFLOPs and GMACs.
	The novel T-FCA and F-FCA mechanisms enhance the model's ability to utilize narrow-band and cross-band information through hardware-friendly operations, thereby more effectively leveraging spatial cues from microphone arrays~\cite{wang2023tf}.

	\subsection{System Outline}
	\label{ssec:System outline}
	
	This subsection provides an overview of the proposed system, detailing the DNN's input, output, and learning target. The input is prepared by first normalizing $X_{m}(t,f)$ using the magnitude mean of the reference channel. Subsequently, the real and imaginary parts of the normalized signal, $\Tilde{X}_m(t,f)$, are stacked to form the input tensor $\mathcal{X} \in \mathbb{R}^{F \times T \times 2M}$. The learning target for the DNN is a complex ideal ratio mask (cIRM) \cite{williamson2015complex}.

	The network outputs a tensor $\mathcal{\hat{Y}} \in \mathbb{R}^{F \times T \times 2}$, which contains the real and imaginary components forming a complex ratio mask. This tensor is designed to approximate the cIRM, denoted as $\mathcal{Y} \in \mathbb{R}^{F \times T \times 2}$, by stacking its real and imaginary parts. The approximation is achieved by minimizing a loss function defined as:
	\allowdisplaybreaks
	\begin{align}
		\mathcal{L} = \alpha\mathcal{L}_{\text{mag}} + (1 - \alpha)\mathcal{L}_{\text{spec}}+\beta\mathcal{L}_{\text{SISDR}},
	\end{align}
	where $\mathcal{L}_{\text{mag}} = \mathrm{MSE}(|\mathcal{Y}|,|\mathcal{\hat{Y}}|)$ and $\mathcal{L}_{\text{spec}} = \mathrm{MSE}(\mathcal{Y},\mathcal{\hat{Y}})$ represent the mean squared error (MSE) losses between the magnitudes of the target and estimated masks, and between the real and imaginary parts~\cite{rong2024gtcrn}, respectively. Here, $|\cdot|$ denotes the magnitude of each element, $\| \cdot \|$ represents the squared norm, and $\alpha$ and $\beta$ are hyperparameters. 
	The scale-invariant SDR loss~\cite{le2019sdr}, denoted as $\mathcal{L}_{\text{SISDR}}$, focuses on maximizing the similarity between the direct-path target speech signal from the reference channel and the estimated speech $\hat{s}$ in the time domain. The estimated speech $\hat{s}$ is derived using $\hat{s} = \mathrm{iSTFT}(\hat{\mathcal{Y}} \odot \mathcal{X}_\mathrm{r})$,
	where $\mathcal{X}_\mathrm{r} \in \mathbb{R}^{F \times T \times 2}$ represents the reference channel of the input, $\odot$ indicates element-wise complex multiplication, and $\mathrm{iSTFT}$ denotes the inverse STFT.
	
	\subsection{Network Architecture}
	\label{ssec:Network Achitecture}
	
	As illustrated in Fig.~\ref{fig:Figure 1.} (a), the proposed LMFCA-Net employs an encoder-decoder architecture with skip connections between the downsampling and upsampling paths. The downsampling path is composed of T-FCA and F-FCA blocks to encode narrow-band and cross-band information, respectively. In the upsampling path, FT-FCA blocks are used to further integrate these features. Although the T-FCA, F-FCA, and FT-FCA blocks share the same basic architecture, they differ in their respective FCA mechanisms. Bottleneck blocks are positioned between the downsampling and upsampling paths, as well as before the final point-wise convolution (PConv). Additionally, depth-wise convolution (DConv) is used in conjunction with PConv throughout the network to reduce computational complexity. The following sections will provide a detailed explanation of the components of LMFCA-Net.
	
	\subsection{Decoupled Fully-Connected Attention (FCA)}
	\label{ssec:FCA}
	
	\begin{figure}[t] \centering
		\includegraphics[width=0.48\textwidth]{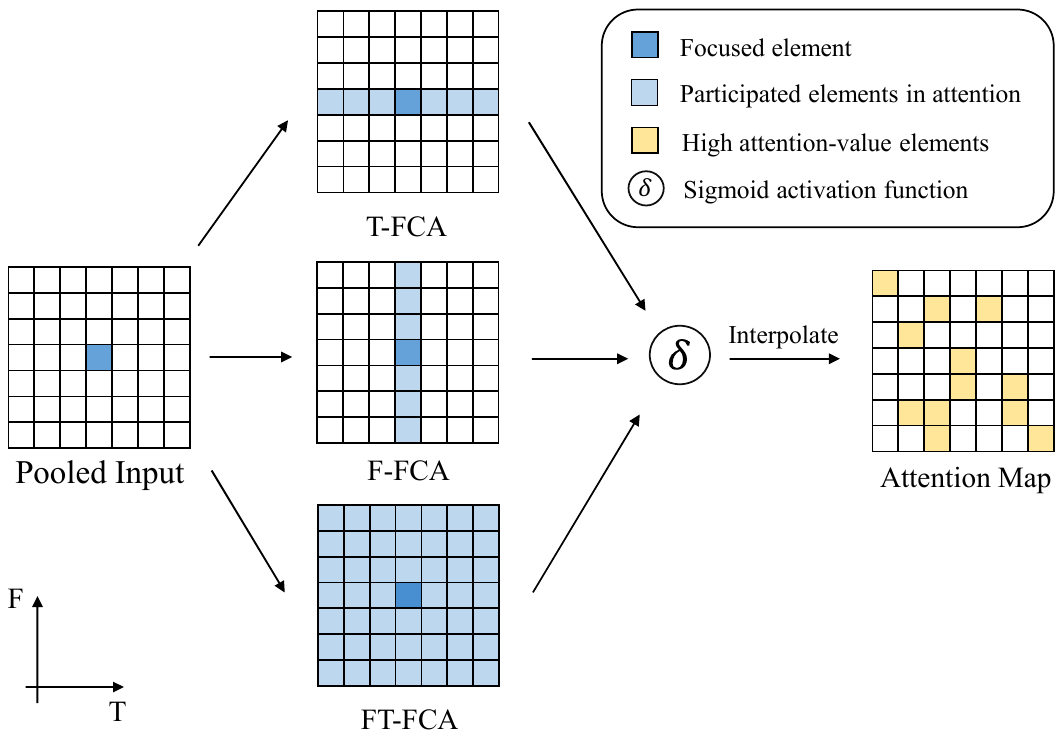}
		\caption{Illustration of FCA mechanisms: T-FCA and F-FCA are designed to model long-range narrow-band and cross-band dependencies, while FT-FCA integrates local features.} \label{fig:fig2}
		\vspace{-0.4cm}
	\end{figure}
	
	As shown in Fig.~\ref{fig:fig2}, the FCA modules take pooled features as their input and generate an attention map to model long-range dependencies. To reduce the computational load of the FCA modules, we apply average pooling with a kernel size of $(2,2)$ on the input features. The concept of T-FCA and F-FCA mechanisms can be expressed as:
	\allowdisplaybreaks
	\begin{align}
		A_{\mathrm{t}}(t, f) = \sum_{t' = 1}^{\hat{T}} W_t(t', f) \odot Z(t', f), \label{eq:tfca}\\ 
		A_{\mathrm{f}}(t, f) = \sum_{f' = 1}^{\hat{F}} W_f(t, f') \odot Z(t, f'), \label{eq:ffca}
	\end{align}
	where $\hat{T}$ and $\hat{F}$ denote the lengths of the time axis and frequency axis of the input feature after pooling, respectively. $A_\mathrm{t}(t, f)$ and $A_\mathrm{f}(t, f)$ represent the attention values for the pooled input feature $Z(t, f)$, and $W_t$ and $W_f$ are the learnable weights. FT-FCA can be simply expressed as:
	\begin{align}
		A_{\mathrm{ft}}(t, f) = \sum_{f' = 1}^{\hat{F}} W_f(t, f') \odot (\sum_{t' = 1}^{\hat{T}} W_t(t', f') \odot Z(t', f')).\label{eq:ftfca}
	\end{align}
	
	Implementing the transformation weights using fully connected layers would result in theoretical complexities of $\mathcal{O}(\hat{F}\hat{T}^2)$, $\mathcal{O}(\hat{F}^2\hat{T})$, and $\mathcal{O}(\hat{F}^2\hat{T} + \hat{F}\hat{T}^2)$ for Eqs.~\eqref{eq:tfca}, \eqref{eq:ffca} and \eqref{eq:ftfca}, respectively, which is suboptimal. However, given the low-rank nature of feature maps~\cite{jaderberg2014speeding}, it is possible to decouple the dense connections of fully-connected layers and implement the transformation weights using convolution \cite{tang2022ghostnetv2}. The implementations can be generally expressed as:
	\begin{align}
		\mathcal{A} = \mathcal{D}_2(\mathcal{D}_1(\mathcal{Z})), \label{eq:implementation}
	\end{align}
	where $\mathcal{D}_1$ and $\mathcal{D}_2$ represent 1D DConvs, and $\mathcal{A}$ and $\mathcal{Z}$ are the attention value tensor and the pooled input feature tensor, respectively. Specifically, $\mathcal{D}_1$ and $\mathcal{D}_2$ are applied along the time-axis for T-FCA, along the frequency-axis for F-FCA, and sequentially along the time-axis followed by the frequency-axis for FT-FCA. This implementation has a theoretical complexity of $\mathcal{O}(K\hat{F} \hat{T})$, where $K$ stands for the kernel size of $\mathcal{D}_1$ and $\mathcal{D}_2$. The resulting attention values are subsequently processed through a Sigmoid activation function and nearest-neighbor upsampling to produce the attention map. Compared to directly applying FT-FCA in all FCA blocks, using T-FCA and F-FCA proves to be more effective, as the generation of the attention map is specifically targeted at narrow-band or cross-band regions. The experimental results in Section.~\ref{sec:results} show that FCA performs well with minimal computational overhead.

        	\begin{table}[t]\centering
		\caption{Performance comparison with open-source multi-channel speech enhancement methods.}
		\label{tab:SOTA comparison}
            \renewcommand{\arraystretch}{1.3} % Adjust row spacing
            \setlength{\tabcolsep}{4pt} % Adjust column spacing
			\begin{tabular}{lccccc}
				\toprule
				\textbf{Method} & WB-PESQ & DNSMOS & GFLOPs & GMACs & RTF  \\
				\midrule
				Noisy & 1.69 & 2.41 & - & - & -\\
				\midrule
				McNet & \textbf{2.79} & \textbf{3.45} & 59.2 & 32.14 & 1.92 \\
                    MC-ConvTasNet & 2.68 & 3.27 & 19.9 & 10.11 & 1.29 \\
				%FaSNet-TAC & 2.47 & 3.29 & 27.9 & 13.77 & 3.26  \\
				LMFCA-Net & 2.51 & 3.30 & \textbf{2.20} & \textbf{1.77} & \textbf{0.16}  \\
				\bottomrule
			\end{tabular}
	\end{table}
    
	\subsection{Sandglass Unit and Bottleneck Block}
	To enhance the efficiency and performance of lightweight neural networks, DConv is often combined with PConv to replace standard convolution in the design, such as MobileNet~\cite{howard2017mobilenets}, to reduce computational cost. DConv captures local information within each channel independently, while PConv enables inter-channel information flow. Building on this, we improve feature capture and network performance by incorporating a sandglass-like structure in our network's Sandglass Unit, using DConv before and after the PConv, inspired by~\cite{zhou2020rethinking}. The specific structure is shown in Fig.~\ref{fig:Figure 1.} (c) and the Bottleneck block is formed by stacking two consecutive Sandglass Units.

	\section{Experimental setup}
	\label{sec:EXPERIMENTS}

	\subsection{Data Preparation}
	\label{ssec:Dataset}
	
	The training samples are generated by convolving clean mono-channel speech with room impulse responses (RIRs). The clean source signals are drawn from the VCTK-DEMAND~\cite{valentini2016investigating} dataset, which includes 11,572 training utterances and 824 testing utterances. The RIRs are generated using the image model~\cite{allen1979image} via Pyroomacoustics~\cite{Scheibler_2018pyroom}, with $200$ rooms and 20 randomly placed sources and microphones per room, resulting in $4,000$ RIRs. The dimensions of the rooms are uniformly sampled, with lengths ranging from $4$ to $10$ meters, widths from $4$ to $10$ meters, and heights from $2.5$ to $3$ meters. The reverberation time, $T_{60}$, is uniformly sampled between $0.3$ and $0.8$ seconds, and the sound source is randomly placed between $0.2$ and $1$ meter from the array's center. 
	The microphone array is modeled as a spherical array consisting of six microphones, each positioned at a radius of $0.1$ meters from the array's center. Noise segments, randomly extracted from the DEMAND dataset, are then added to the convolved speech to create the final noisy dataset. Noisy utterances are generated with a signal-to-noise ratio (SNR) randomly chosen within the range of $[0, 12]$ dB. Test and validation sets are generated once and reused across all experiments to ensure replicability. We also evaluate our model’s generalization with a blind test on the CHIME-3 real-world dataset~\cite{barker2015third} \footnote{Audio demos and code can be found at \href{https://lmfcanet.github.io/}{https://lmfcanet.github.io/.}}.

	\subsection{Configurations}
	\label{ssec:Configurations}
	
	All audio files are sampled at $16$ kHz, with channel five serving as the reference for all experiments. The STFT is applied using a $510$-sample Hanning window and a hop length of $255$ samples, ensuring that the number of frequency bins is divisible by $8$. The audio files are segmented into approximately $3$-second clips, padded to make the frame count divisible by $8$. Downsampling and upsampling are performed using max pooling and transposed convolution with a kernel size of $(2,2)$. 
	The model's output channels are configured as $C1=48$, $C2=96$, $C3=224$, and $C4=480$. The kernel size of the 2D DConv in the Sandglass Units is set to $(3,3)$, and the 1D DConv in FCAs has a kernel size of $5$. In both the FCA Block and Sandglass Unit, the first PConv reduces the channel dimension to half of the output channel number, which is then expanded to the full output channel number by DConv or PConv. $\alpha$ and $\beta$ in the loss function are set to $0.1$ and $10^{-4}$, respectively. The model is trained with the Adam optimizer, with an initial learning rate of $10^{-4}$. The learning rate is halved if the performance on the validation set does not improve for five consecutive epochs. The batch size is set to four, and the model is trained on an NVIDIA Tesla V100 GPU.
	
	The evaluation metrics include WB-PESQ~\cite{rix2001perceptual} and DNSMOS P.808 (DNSMOS)~\cite{reddy2021dnsmos} for assessing speech quality, as well as MACs, floating-point operations per second (FLOPs), and offline real-time factor (RTF) to evaluate computational complexity. WB-PESQ is measured using $s(t)$ as the target signal. The offline RTF is calculated on an Intel Core i5-1135G7 @ 2.40GHz CPU by dividing the processing time by the input duration, and an RTF lower than one indicates that the system processes the input faster than real-time.
	
	\section{Results and discussions}
	\label{sec:results}
	\subsection{Performance Comparison}
	\label{sssec:Performance Comparison}
    
% \begin{table*}[h!]
% \setlength\tabcolsep{3pt}
% \caption{Performance comparison with lightweight models. Mono-channel stands for using only the reference channel of the input.}
% \label{tab:combined}
% \centering
% \renewcommand{\arraystretch}{1.3} % Adjust row spacing
% \setlength{\tabcolsep}{20pt} % Adjust column spacing
% \begin{tabular}{lccccc}
% \toprule
% \textbf{Method}  & WB-PESQ & DNSMOS & GFLOPs & GMACs  & RTF \\
% \midrule
% Noisy  & 1.69 & 2.41 & - & - & -\\
% \midrule
% \multicolumn{6}{c}{\textbf{Multi-channel}} \\
% \midrule
% Wave-U-Net   & 2.23 &  3.16 & 4.83 & 2.42 & 0.19\\
% GTCRN & 2.19 & 2.96 & \textbf{0.14} & \textbf{0.06} & \textbf{0.07} \\
% LMFCA-Net & \textbf{2.51} & \textbf{3.30} & 2.20 & 1.77 & 0.16\\
% \midrule
% \multicolumn{6}{c}{\textbf{Mono-channel}} \\
% \midrule
% Wave-U-Net   & 2.15 & 3.12 & 4.72 & 2.39 & 0.18\\
% GTCRN & 2.09 & 2.91 & \textbf{0.12} & \textbf{0.04} & \textbf{0.06}\\
% LMFCA-Net & \textbf{2.24} & \textbf{3.22} & 2.16 & 1.75 & 0.16\\
% \bottomrule
% \end{tabular}%

% \end{table*}
	
	We first evaluate our model by comparing it with McNet~\cite{yang2023MCNET} and MC-ConvTasNet~\cite{zhang2020mcconvtas}, both of which are designed for multi-channel speech enhancement. McNet leverages recurrent units as its backbone, while MC-ConvTasNet operates in the time domain and introduces the use of 2D convolution to capture inter-channel spatial information. We use these two open-source models without any modifications and train them to convergence on our dataset. As shown in Table~\ref{tab:SOTA comparison}, the proposed LMFCA-Net is significantly lighter and faster, as evidenced by its lower GFLOPs, GMACs, and RTF, while maintaining performance comparable to these models.

	Next, we compare our model with other lightweight models. Given the limited research on designing lightweight models specifically for multi-channel speech processing, we adapted Wave-U-Net~\cite{stoller2018wave} and GTCRN~\cite{rong2024gtcrn} into multi-channel models by simply increasing the number of input channels, following~\cite{stoller2018wave}. The former was originally designed for speech separation, while the latter is a more recent lightweight speech enhancement model based on the U-Net architecture. Table~\ref{tab:Baseline Comparison} shows that LMFCA-Net strikes a good balance between efficiency and quality, as demonstrated by its superior performance in WB-PESQ and DNSMOS, while maintaining an RTF similar to other lightweight models.

	Finally, we compared all lightweight baseline models under mono-channel conditions by training the mono-channel version of LMFCA-Net and other baselines using the reference channel of the input. The results in Table~\ref{tab:mono} indicate that LMFCA-Net is more effective in utilizing spatial cues from multiple microphones. This is evidenced by the WB-PESQ improvement of over 0.25 when transitioning from mono-channel to multi-channel input for LMFCA-Net, compared to only about 0.1 for other lightweight models.

	\begin{table}[t]
		\centering
		\caption{Performance comparison with lightweight models (multi-channel).}
            \renewcommand{\arraystretch}{1.3} % Adjust row spacing
            \setlength{\tabcolsep}{4pt} % Adjust column spacing
		\label{tab:Baseline Comparison}
			\begin{tabular}{lccccc}
				\toprule
				\textbf{Method}  & WB-PESQ & DNSMOS & GFLOPs & GMACs  & RTF \\
				\midrule
				Noisy  & 1.69 & 2.41 & - & - & -\\
				\midrule
				Wave-U-Net   & 2.23 &  3.16 & 4.83 & 2.42 & 0.19\\
				GTCRN & 2.19 & 2.96 & \textbf{0.14} & \textbf{0.06} & \textbf{0.07} \\
				LMFCA-Net & \textbf{2.51} & \textbf{3.30} & 2.20 & 1.77 & 0.16\\
				\bottomrule
			\end{tabular}
	\end{table}

	\begin{table}[t]
		\centering
		\caption{Performance comparison with lightweight models (mono-channel).}
		\label{tab:mono}
            \renewcommand{\arraystretch}{1.3} % Adjust row spacing
            \setlength{\tabcolsep}{4pt} % Adjust column spacing
			\begin{tabular}{lccccc}
				\toprule
				\textbf{Method}  & WB-PESQ & DNSMOS & GFLOPs & GMACs  & RTF \\
				\midrule
				Noisy  & 1.69 & 2.41 & - & - & -\\
				\midrule
				Wave-U-Net   & 2.15 & 3.12 & 4.72 & 2.39 & 0.18\\
				GTCRN & 2.09 & 2.91 & \textbf{0.12} & \textbf{0.04} & \textbf{0.06}\\
				LMFCA-Net & \textbf{2.24} & \textbf{3.22} & 2.16 & 1.75 & 0.16\\
				\bottomrule
			\end{tabular}
	\end{table}

	\subsection{Ablation Study}
	\label{ssec:Ablation_study}

	We assess the impact of the components by removing or replacing certain modules within the model. In examining the FCA mechanism within T-FCA, F-FCA, and FT-FCA blocks, we removed the FCA branch to study its effect. Results in Table~\ref{tab:Ablation Study} show that FCA yields a significant performance gain while adding only minimal computation and latency. The Sandglass Unit also serves as a crucial building block, encoding richer local information than PConv. Additionally, we examined the effectiveness of the two-stage framework by replacing T-FCA and F-FCA modules with FT-FCA. The experimental results demonstrate that the two-stage strategy yields better performance than using FT-FCA alone.
	
	\begin{table}[h]
		\centering
		\caption{Ablation study results.}
		\label{tab:Ablation Study}
            \renewcommand{\arraystretch}{1.3} % Adjust row spacing
            \setlength{\tabcolsep}{4pt} % Adjust column spacing
			\begin{tabular}{lcccc}
				\toprule
				\textbf{Method} & WB-PESQ & DNSMOS  & GFLOPs & RTF \\
				\midrule
				Noisy & 1.69 & 2.41 & - & -  \\
				\midrule
				LMFCA-Net & 2.51 & 3.30 & 2.20 & 0.16 \\
				w/o FCA & 2.41 & 3.19 & 1.88 & 0.15 \\
				\makecell[l]{replace Sandglass Unit \\\ \ \ with PConv} & 2.46  & 3.27 & 2.21 & 0.16 \\
				\makecell[l]{replace T-FCA \& F-FCA\\ with FT-FCA} & 2.48 & 3.26 & 2.20 & 0.16 \\
				\bottomrule
			\end{tabular}
	\end{table}

	\section{Conclusion}
	\label{sec:CONCLUSION}
	
	This paper presented LMFCA-Net, a lightweight multi-channel speech enhancement model with novel T-FCA and F-FCA mechanisms to capture sub-band and cross-band information through hardware-friendly operations. This approach focus on addressing the need for efficient DNN-based microphone array processing in resource-constrained environments. 
    Experimental results confirm the effectiveness of the two-stage T-FCA and F-FCA modeling, as well as LMFCA-Net, demonstrating its strong performance with low computational load and latency.
    Our work is presented in an offline setting, but with its low RTF, LMFCA-Net has significant potential for adaptation to online processing through simple techniques like buffering, as demonstrated in our source code. We also plan to further reduce its algorithmic latency for online applications by implementing causal convolution and attention units in the future.

	\footnotesize
	\bibliographystyle{ieeetr}
	\bibliography{Bib_MABFSE, refs}

\end{document}